\documentclass{aa}

\def\vec#1{\ensuremath{\mathchoice{\mbox{\boldmath$\displaystyle#1$}}
{\mbox{\boldmath$\textstyle#1$}}
{\mbox{\boldmath$\scriptstyle#1$}}
{\mbox{\boldmath$\scriptscriptstyle#1$}}}}

\usepackage[varg]{txfonts}

\begin{document}

\title{Impact of basic angle variations on the parallax zero point for a scanning astrometric satellite}

\titlerunning{Basic angle variations and parallax zero point}

\author{Alexey G. Butkevich\inst{1,2} \and Sergei A. Klioner\inst{2} \and Lennart Lindegren\inst{3} \and David Hobbs\inst{3} \and Floor van Leeuwen\inst{4}}

\authorrunning{A.G. Butkevich et al.}

\institute{
         Pulkovo Observatory, Pulkovskoye shosse 65, 196140 Saint-Petersburg, Russia\\
           \email{ag.butkevich@gmail.com}
         \and
           Lohrmann Observatory, Technische Universit\"at Dresden,
                01062 Dresden, Germany\\
           \email{sergei.klioner@tu-dresden.de}
         \and
		Lund Observatory, Department of Astronomy and Theoretical Physics, Lund University, Box 43, 22100 Lund, Sweden\\
		   \email{[lennart; david]@astro.lu.se}
         \and
         Institute of Astronomy, Madingley Road, Cambridge CB3 OHA, UK\\
		   \email{fvl@ast.cam.ac.uk}}

\date{Received; accepted}

\abstract
{Determination of absolute parallaxes by means of a scanning astrometric satellite
such as {\sc Hipparcos} or \emph{Gaia} relies on the short-term stability of the so-called basic angle between
the two viewing directions. Uncalibrated variations of the basic angle may produce systematic
errors in the computed parallaxes.}
{We examine the coupling between a global parallax shift and specific variations of the
  basic angle, namely those related to the satellite attitude with respect to the Sun.}
{The changes in observables produced by small perturbations of the basic angle, attitude, and
parallaxes are calculated analytically. We then look for a combination of perturbations that has
no net effect on the observables.}
{In the approximation of infinitely small fields of view, it is shown that certain perturbations
of the basic angle are observationally indistinguishable from a global shift of the parallaxes.
If such perturbations exist, they cannot be calibrated
from the astrometric observations but will produce a global parallax bias. Numerical simulations
of the astrometric solution, using both direct and iterative methods, confirm this theoretical
result. For a given amplitude of the basic angle perturbation, the parallax bias is smaller for
a larger basic angle and a larger solar aspect angle. In both these respects \emph{Gaia} has a more
favourable geometry than {\sc Hipparcos}. In the case of \emph{Gaia}, internal metrology is used to monitor basic angle variations.
Additionally, Gaia has the advantage of detecting numerous quasars, which can be used to verify the parallax zero point.}
{}


\maketitle

\section{Introduction}

The European Space Agency's space astrometry mission \emph{Gaia},
aiming to determine astrometric parameters for at least one billion stars
with accuracies reaching 10~microarcseconds \citep{2012Ap&SS.341...31D},
was launched in December 2013 \citep{2016A&A...595A...1G}.
\emph{Gaia} is based on similar observation principles as
the highly successful pioneering astrometric mission {\sc Hipparcos}
\citep{1997ESASP1200.....E}. In particular, both satellites use an optical system
providing two viewing directions separated by a wide angle, referred to as the basic angle
\citep{2001A&A...369..339P}. The goal of the present paper is to show
how certain time-dependent variations of the basic angle can bias the
parallax zero point of an astrometric solution derived from
observations by such a scanning astrometric satellite.

The data processing for an astrometric satellite should, as far as possible,
be based on the principle of self-calibration \citep{2011EAS....45..109L}.
This means that the same observational data are used to determine both the
scientifically interesting astrometric parameters and the so-called ``nuisance parameters''
that describe the instrument calibration, satellite attitude, and other relevant
parts of the observation model
\citep{2012A&A...538A..78L}. The self-calibration is, however, of
limited applicability when variations of different parameters do not produce
fully independent effects in the observables. Such situations can occur when
the variation of certain parameters leads to changes in the observables that
resemble the changes produced by the variation of some other parameters.
The more similar the changes in the observables are, the stronger the correlation
between the parameters. If the changes are identical, the problem of
parameter estimation is degenerate: the same set of observables is equally
well described by different sets of parameter values.

As long as the degeneracy involves only nuisance parameters, it has no
effect on the astrometric solution and only leads to an arbitrary but
unimportant shift of the respective nuisance parameters.
However, if there is a degeneracy between the astrometric and nuisance
parameters, the self-calibration process will in general lead to biased astrometry.
The celestial reference frame is an example of a complete degeneracy between
the astrometric and attitude parameters, which can only be lifted by means
of external data, in this case the positions and proper motions of a number
of extragalactic objects \citep{1997A&A...323..620K,2012A&A...538A..78L}.
Concerning the instrument calibration parameters, it is possible to formulate
the calibration model in such a way that (near-)degeneracies are avoided among
its parameters, as well as between the calibration and attitude parameters.

However, it is still possible that the actual physical variations of the instrument
contain components that are degenerate with the astrometric parameters.
By definition, such variations cannot be detected internally by the astrometric
solution (e.g., from an analysis of the residuals), but only through a comparison
with external data, e.g.\ astrophysical information or independent metrology.
In Gaia the latter is chosen as will be detailed in Section~\ref{sec:relevance}.

It turns out that the basic angle is an important example of a quantity
that could vary in a way that cannot be fully calibrated from observations.
Already in the early years of the {\sc Hipparcos} project it was realised
that certain periodic variations of the basic angle, caused by a non-uniform
heating of the satellite by the solar radiation, lead to a global shift of the
parallaxes \citep{lindegren77}.
Subsequent analyses \citep{1995A&A...304...52A,2005A&A...439..805V}
concluded that the possible effect on the {\sc Hipparcos} parallaxes
was negligible, suggesting a very good short-term stability of the
basic angle in that satellite.

For \emph{Gaia} the situation is different. The much higher accuracy targeted
by this mission necessitates a very careful consideration of possible biases
introduced by uncalibrated instrumental effects, including basic angle variations.
This is even more evident in view of the very significant ($\sim$1~mas amplitude)
basic angle variations measured by the on-board metrology system of \emph{Gaia}
\citep{2016A&A...595A...1G,2016A&A...595A...4L}. In this context the
near-degeneracy between a global parallax zero point error and a possible
basic-angle variation induced by solar radiation is particularly relevant. The
theoretical analysis of the problem presented here expands and clarifies earlier analytical
results by \citet{lindegren77,LL:GAIA-LL-057} and \citet{2005A&A...439..805V}.

An analytical treatment of the problem is given in Sect.~\ref{s:derivations}.
Section~\ref{s:simulations} presents the results of numerical experiments
confirming the theoretical expectations. In Sect.~\ref{sec:disc} we consider the
practical implications of results. Some concluding remarks are given in
Sect.~\ref{s:conclusion}.

\section{Theory}
\label{s:derivations}

In this section we consider how small perturbations of various parameters
change the observed quantities.  We first demonstrate that, to first
order in the small angles, arbitrary variations of observables are equivalent
to certain variations of the basic angle and attitude
(Sects.~\ref{sec-reference-system}--\ref{sec-attitude}). Then we find the
changes of observables due to a common shift of all parallaxes
(Sect.~\ref{sec-parallax}). Combining these results, we derive in
Sects.~\ref{sec-parallax-ba-attitude-general}--\ref{sec-parallax-ba-attitude-harmonic}
the specific variations of the basic angle and attitude that are observationally
indistinguishable from a common shift of the parallaxes.

\subsection{Reference system}
\label{sec-reference-system}

To study the coupling between the instrument parameters and parallax, it is convenient to make use of the rotating reference system aligned with the fields of view. This system, known as the Scanning Reference System (SRS) in the \emph{Gaia} nomenclature \citep{2012A&A...538A..78L}, is represented by the instrument axes $\vec{x}$, $\vec{y}$, $\vec{z}$ (Fig.~\ref{fig:system}), with $\vec{z}$ directed along the nominal spin axis of the satellite, $\vec{x}$ bisecting the two viewing directions separated by the basic angle $\Gamma$, and $\vec{y}=\vec{z}\times\vec{x}$.  The direction towards an object is specified by the unit vector
\begin{equation}\label{u_pqr}
\vec{u}=\vec{x}\cos\varphi\cos h+\vec{y}\sin\varphi\cos h+\vec{z}\sin h\,,
\end{equation}
with the instrument angles $\varphi$
and $h$ describing the position of the object with respect to the SRS
(Fig.~\ref{fig:system}).
For a star in the preceding field of view (PFoV) $\varphi\simeq +\Gamma/2$,
while in the following field of view (FFoV) $\varphi\simeq -\Gamma/2$.

\subsection{Field angles}

\begin{figure}
  \resizebox{\hsize}{!}{\includegraphics{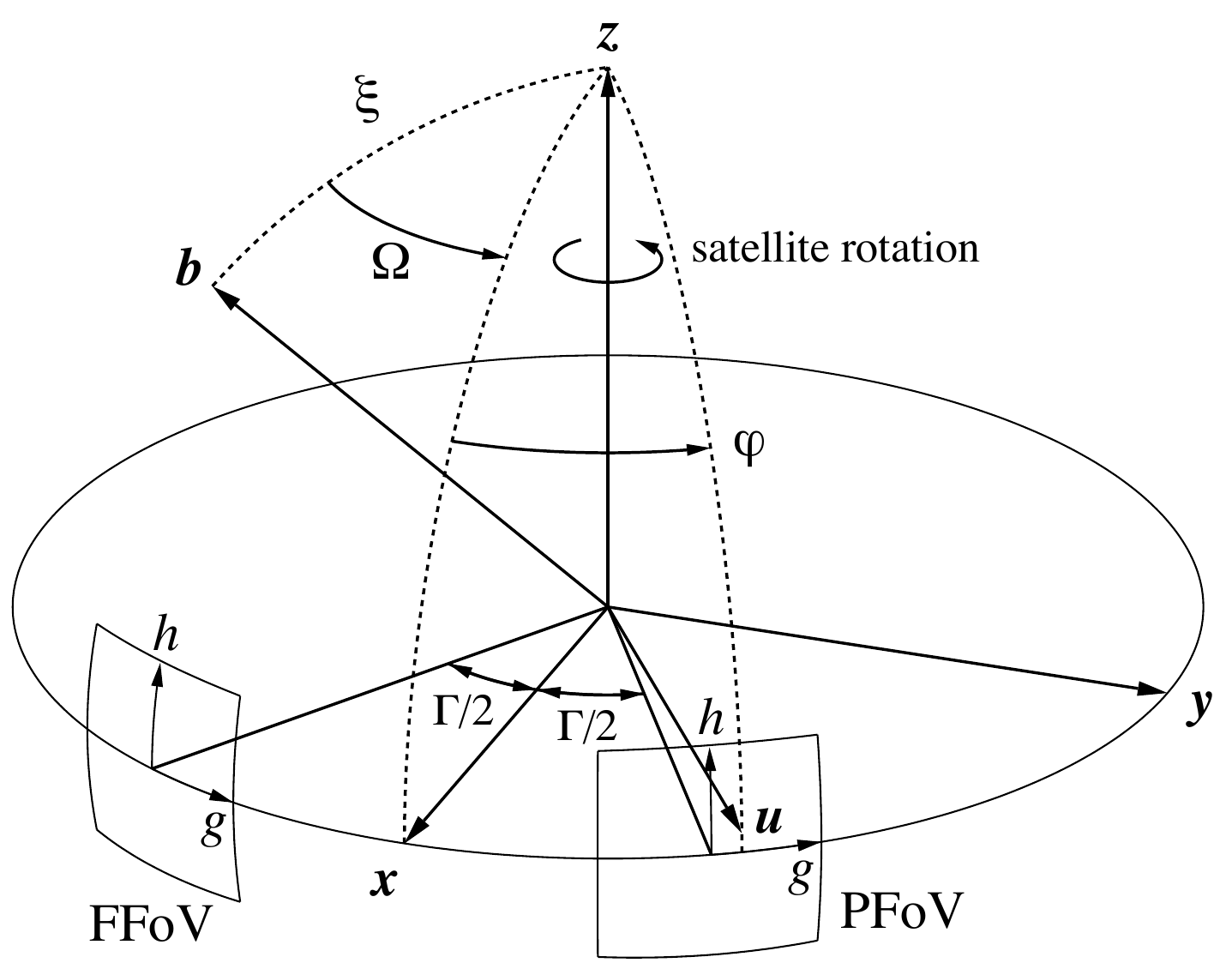}}
  \caption{Definition of the instrument axes $\vec{x}$, $\vec{y}$, $\vec{z}$ of the
  Scanning Reference System (SRS), the basic angle $\Gamma$, and the field angles
  $g$ and $h$ specifying the observed direction to a star ($\vec{u}$) in either
  field of view. $\varphi$ is the along-scan instrument angle of the star. In the SRS
  the direction to the solar system barycentre, $\vec{b}$, is specified by the angles
  $\xi$ and $\Omega$.}
  \label{fig:system}
\end{figure}

An observation consists of a measurement of the coordinates of a
stellar image in the focal plane at a particular time. In practice the
measurement is expressed in detector coordinates (e.g.\ pixels), but
we consider here an idealised system providing a direct measurement of
the two field angles $g$ and $h$ in the relevant field of view. While
the across-scan field angle $h$ coincides with the corresponding
instrument angle, the along-scan field angle $g$ is reckoned from
the centre of the corresponding field of view in the direction of the
satellite rotation (Fig.~\ref{fig:system}). Projected on the sky, the
field-of-view centre defines two viewing directions separated by the
basic angle $\Gamma$. Thus,
\begin{equation}\label{eq:gpf}
\left.
\begin{aligned}
g_\mathrm{p}&=\varphi-\Gamma/2\quad
\text{in the preceding field of view}\\
g_\mathrm{f}&=\varphi+\Gamma/2\quad
\text{in the following field of view}
\quad\end{aligned}
\right\}\,,
\end{equation}
where subscripts p and f denote values for the
preceding and following field of view, respectively.  We assume
that the instrument is ideal except for the basic angle $\Gamma$,
which can deviate from its nominal (conventional) value
$\Gamma_\mathrm{c}$ by a time-dependent variation:
\begin{equation}\label{eq:g}
\Gamma(t)=\Gamma_\mathrm{c}+\delta\Gamma(t) \, .
\end{equation}
It is important to note that the along-scan field angle $g$, as defined
here, is not the same as the along-scan field angle $\eta$ normally used
in the context of the \emph{Gaia} data processing \citep{2012A&A...538A..78L}.
While $\eta$ is measured from a fixed, conventional origin at
$\varphi=\pm\Gamma_\mathrm{c}/2$, our $g$ is measured from the
actual, variable field centre at $\varphi=\pm\Gamma(t)/2$.
This difference
motivates the change in notation from $\eta$ to $g$. For consistency,
a corresponding change is made in the across-scan direction, although our
$h$ is the same as the across-scan field angle $\zeta$ used in the \emph{Gaia}
data processing.

\subsection{Variations of the field angles due to a change in the basic angle}
\label{ss_basangle}

Any increase or decrease of the basic angle makes the fields of view
move further from each other or closer together. This, in turn, changes the
observed field angle $g$ for a given stellar image. However, since the
attitude (celestial pointing of the SRS axes) is unchanged, the value
of $\varphi$ for a given star is not affected by the basic angle. For
example, let us consider the preceding field of view. An increase of
the basic angle causes the observed image to be shifted with respect
to the centre of the field of view so that the observed along-scan
field angle $g_\mathrm{p}$ is decreased. The opposite effect takes
place in the following field of view. The across-scan field angles
$h_\mathrm p$ and $h_\mathrm f$ are obviously not affected.
The variations of the field angles caused by the basic-angle variation
$\delta\Gamma$ are therefore
\begin{equation}\label{eq:alpf_gamma}
\left.
\begin{aligned}
\delta g_\mathrm{p}&=-{\textstyle\frac{1}{2}}\,\delta\Gamma\\
\delta g_\mathrm{f}&=+{\textstyle\frac{1}{2}}\,\delta\Gamma\\
\delta h_\mathrm{p}&=0\\
\delta h_\mathrm{f}&=0\\
\end{aligned}
\quad\right\}\,.
\end{equation}
\noindent
This agrees with Eq.~(\ref{eq:gpf}) taking into account that $\delta\varphi=0$.

\subsection{Variations of the field angles due to a change in the attitude}
\label{sec-attitude}

A quaternion representation is used to parametrise the attitude of
\emph{Gaia} \cite[Appendix A]{2012A&A...538A..78L}. Here it is more
convenient to describe small changes in the attitude by means of three
small angles $\delta_x$, $\delta_y$, and $\delta_z$ representing the
rotations around the corresponding SRS axes. Since the direction $\vec{u}$ to the star
is regarded here as fixed, the corresponding changes in the observed
field angles are found to be
\begin{equation}\label{eq:alac_att}
\left.
\begin{aligned}
\delta g_\mathrm{p}&=-\delta_z\\
\delta g_\mathrm{f}&=-\delta_z\\
\delta h_\mathrm{p}&=\cos(\Gamma_\mathrm{c}/2)\,\delta_y-\sin(\Gamma_\mathrm{c}/2)\,\delta_x\\
\delta h_\mathrm{f}&=\cos(\Gamma_\mathrm{c}/2)\,\delta_y+\sin(\Gamma_\mathrm{c}/2)\,\delta_x
\end{aligned}
\quad\right\}\,.
\end{equation}
In these and following equations, we neglect terms of the second and
higher orders in $\delta_x$, $\delta_y$, $\delta_z$, and $\delta\Gamma$.
To this approximation we can use $\Gamma_\mathrm{c}$ instead of $\Gamma$
in the trigonometric factors.

\subsection{Combined changes in the basic angle and attitude}
\label{sec-combined}

Combining Eqs.~(\ref{eq:alpf_gamma}) and (\ref{eq:alac_att}) we see that
a simultaneous change of the basic angle by $\delta\Gamma$ and of the
attitude by $\delta_x$, $\delta_y$, $\delta_z$ gives the following total
change of the field angles:
\begin{equation}\label{eq:gamma_att}
\left.
\begin{aligned}
\delta g_\mathrm{p}&=-{\textstyle\frac{1}{2}}\,\delta\Gamma-\delta_z\\
\delta g_\mathrm{f}&=+{\textstyle\frac{1}{2}}\,\delta\Gamma-\delta_z\\
\delta h_\mathrm{p}&=\cos(\Gamma_\mathrm{c}/2)\,\delta_y-\sin(\Gamma_\mathrm{c}/2)\,\delta_x\\
\delta h_\mathrm{f}&=\cos(\Gamma_\mathrm{c}/2)\,\delta_y+\sin(\Gamma_\mathrm{c}/2)\,\delta_x
\end{aligned}
\quad\right\}\,.
\end{equation}
An exact inversion of this system of equations gives
\begin{equation}\label{eq:gamma_att_inv}
\left.
\begin{aligned}
\delta_x&=\frac{1}{2\sin\left(\Gamma_\mathrm{c}/2\right)}\left(\delta h_\mathrm{f}
-\delta h_\mathrm{p}\right)\\
\delta_y&=\frac{1}{2\cos(\Gamma_\mathrm{c}/2)}\left(\delta h_\mathrm{p}
+\delta h_\mathrm{f}\right)\\
\delta_z&=-\frac{1}{2}\left(\delta g_\mathrm{p}+\delta g_\mathrm{f}\right)\\
\delta\Gamma&=\delta g_\mathrm{f}-\delta g_\mathrm{p}
\end{aligned}
\quad\right\}\,.
\end{equation}
The first two equations in (\ref{eq:gamma_att_inv}) show that arbitrary small
changes in the across-scan field angles $\delta h_\mathrm{p}$ and
$\delta h_\mathrm{f}$ to first order can be represented as changes in the
attitude by $\delta_x$ and $\delta_y$. Similarly, the last two equations
show that arbitrary changes in the along-scan field angles $\delta g_\mathrm{p}$
and $\delta g_\mathrm{f}$ can be represented as a combination of a change of
the basic angle $\delta\Gamma$ and a change in the attitude by $\delta_z$.

In general, an arbitrary perturbation of the observed stellar positions, being a smooth function
of time and stellar position, clearly result in a smooth, time-dependent variation of
$\delta g_\mathrm{p}$, $\delta g_\mathrm{f}$, $\delta h_\mathrm{p}$,
and $\delta g_\mathrm{f}$. From Eq.~(\ref{eq:gamma_att_inv}) it follows that such
a perturbation is observationally indistinguishable from a certain time-dependent
variation of $\delta_x$, $\delta_y$, $\delta_z$, and $\delta\Gamma$.

\subsection{Variations of the field angles due to a change in the parallax}
\label{sec-parallax}

The position of the barycentre of the solar system with respect to the
instrument can be specified by a distance $R$ (in au) and two angular coordinates.
We take the angular coordinates to be $\xi$ and $\Omega$ defined as in Fig.~\ref{fig:system}.
According to the scanning law, $\xi$ is nearly constant while $\Omega$ is
increasing with time as the satellite spins.
The barycentric position of the satellite is
\begin{equation}\label{R-vector}
\vec{R}=R\left(-\vec{x}\cos\Omega\sin\xi+\vec{y}\sin\Omega\sin\xi-\vec{z}\cos\xi\right)\,.
\end{equation}
The observed direction $\vec{u}$ to a star is given by Eq.~(4) of
\citet{2012A&A...538A..78L} as a function of the astrometric parameters
of the star. Linearisation yields the
change in the direction caused by a small change of the parallax
$\delta\varpi$:
\begin{equation}\label{delta_u}
    \delta\vec{u}=\vec{u}\times\left(\vec{u}\times\vec{R}\,\delta\varpi\right)\,.
\end{equation}

We now assume that the direction to a star is changed only from
a change of its parallax, while the basic angle and attitude are kept
constant. The fixed basic angle implies $\delta\varphi=\delta g$.
The fixed attitude means that $\vec{x}$, $\vec{y}$, and $\vec{z}$
are constant, so that Eq.~(\ref{u_pqr}) gives the change in direction
\begin{equation}\label{eq:du}
\delta\vec{u} = \vec{v}\cos h\,\delta g + \vec{w}\,\delta h\,,
\end{equation}
where
\begin{equation}\label{eq:vw}
\left.\begin{aligned}
\vec{v}&=
-\vec{x}\sin\varphi+\vec{y}\cos\varphi\\
\vec{w}&=
-\vec{x}\cos\varphi\sin h-\vec{y}\sin\varphi\sin h+\vec{z}\cos h
\end{aligned}
\quad\right\}
\end{equation}
are unit vectors in the directions of increasing $\varphi$ and $h$, respectively.
They are evidently orthogonal to each other and to $\vec{u}$. Equating
$\delta\vec{u}$ from (\ref{delta_u}) and (\ref{eq:du}) and taking the
dot product with $\vec{v}$ and $\vec{w}$ gives
\begin{equation}\label{eq:eta_pi}
\left.
\begin{aligned}
\cos h\,\delta g&=-\vec{v}^\prime\vec{R}\,\delta\varpi\\
\delta h&=-\vec{w}^\prime\vec{R}\,\delta\varpi
\end{aligned}
\quad\right\}\,.
\end{equation}
Substituting Eqs.~(\ref{R-vector}) and (\ref{eq:vw}) we find
\begin{equation}\label{eq:alac_par}
\left.
\begin{aligned}
\cos h\,\delta g&=-\sin\left(\Omega+\varphi\right)\sin\xi\, R\,\delta\varpi\\
\delta h&=\left[\cos h\cos\xi-\cos(\Omega+\varphi)\sin h\sin\xi\right]\,R\,\delta\varpi
\end{aligned}
\quad\right\}\,.
\end{equation}

Up to this point the derived formulae are valid throughout the field of view to first
order in the (very small) variations denoted with a $\delta$. To proceed, we now
consider  a star observed at the center of either field of view, so that $g=h=0$ and
$\varphi=\pm\,\Gamma_\mathrm{c}/2$.  In this case the variations of
the field angles caused by $\delta\varpi$ become
\begin{equation}\label{eq:gh_par}
\left.
\begin{aligned}
\delta g_\mathrm{p}&=-\sin\left(\Omega+\Gamma_\mathrm{c}/2\right)\sin\xi\, R\,\delta\varpi\\
\delta g_\mathrm{f}&=-\sin\left(\Omega-\Gamma_\mathrm{c}/2\right)\sin\xi\, R\,\delta\varpi\\
\delta h_\mathrm{p}&=\cos\xi\, R\,\delta\varpi\\
\delta h_\mathrm{f}&=\cos\xi\, R\,\delta\varpi\,.
\end{aligned}
\quad\right\}\,.
\end{equation}
Considering only stars at the centre of either field of view effectively means that
we neglect the finite size of the field of view. In both {\sc Hipparcos} and \emph{Gaia}
the half-size of the field of view is $\Phi < 10^{-2}$~rad. Since $|g|$, $|h|<\Phi$,
neglected terms in Eq.~(\ref{eq:gh_par}) are of the order of $\Phi\times\delta$,
where $\delta$ represents any of the quantities $\delta\Gamma$, $\delta_x$, etc.
Equation~(\ref{eq:gh_par}) is therefore expected to be accurate to $<1$\% at any point
in the field of view. The implications of this approximation are further discussed in
Sect.~\ref{sec:fov}.

\subsection{Relation between the changes in parallax, basic angle, and attitude}
\label{sec-parallax-ba-attitude-general}

Substituting Eq.~(\ref{eq:gh_par}) into Eq.~(\ref{eq:gamma_att_inv}) we readily
obtain a relation between the change in parallax and the corresponding changes
in basic angle and attitude:
\begin{equation}\label{eq:delta_par}
\left.
\begin{aligned}
\delta_x&=0\\
\delta_y&=\cos\xi\sec(\Gamma_\mathrm{c}/2)\,R\,\delta\varpi\\
\delta_z&=\sin\Omega\sin\xi\cos(\Gamma_\mathrm{c}/2)\,R\,\delta\varpi\\
\delta\Gamma&=2\cos\Omega\sin\xi\sin(\Gamma_\mathrm{c}/2)\,R\,\delta\varpi
\end{aligned}
\quad\right\}\,.
\end{equation}
These equations should be interpreted as follows: a change in
parallax by $\delta\varpi$ is observationally indistinguishable (to order
$\Phi\times\delta$) from a simultaneous change of the attitude by $\delta_x$,
$\delta_y$, $\delta_z$ and of the basic angle by $\delta\Gamma$.
The formulae were derived for one star, but if $\delta\varpi$ is the same
for all stars, they hold for all observations of all stars.
Equation~(\ref{eq:delta_par}) therefore defines the specific variations of
the attitude and basic angle that mimic a global shift in parallax.
Remarkably, the rotation around the $\vec{x}$ axis is not affected by the
global parallax change, while the rotation around the $\vec{y}$ axis is
independent of $\Omega$ and therefore, in practice, almost constant.

In a global astrometric solution all the attitude and stellar parameters
(including $\varpi$) are simultaneously fitted to the observations of $g$
and $h$. A specific variation in the basic angle of the form
$\delta\Gamma(t)\propto\cos\Omega\sin\xi\,R$ will then lead to a global
shift of the fitted parallaxes (together with some time-dependent attitude
errors $\delta_y$, $\delta_z$). Since the effects of such a basic-angle variation
are fully absorbed by the attitude parameters and parallaxes, the variation is
completely degenerate with the stellar and attitude model and cannot be
detected from the residuals of the fit.

For a satellite in orbit around the Earth (as {\sc Hipparcos}) or near L$_2$
(as \emph{Gaia}), $R$ is approximately constant. In order to achieve a stable
thermal regime of the instrument for a scanning astrometric satellite one
typically chooses a scanning law with a constant angle between
the direction to the Sun and the spin axis -- the so-called solar
aspect angle. This means that angle $\xi$ is nearly constant as well
(see Sect.~\ref{sec:helio}).
In the next section we consider the case when $R$ and
$\xi$ are exactly constant. Nevertheless, in reality both $R$ and $\xi$ are
somewhat time-dependent, and this case is discussed in Sect.~\ref{s:time}.

Since $\xi$ and $R$ are nearly constant, the degenerate component of the
basic-angle variation is essentially of the form $\cos\Omega$, which is
periodic with the satellite spin period relative to the solar-system barycentre.
This leads to a fundamental design requirement for a scanning astrometry
satellite, namely that the basic angle should not have significant periodic
variations with a period close to the period of rotation of the satellite, and
especially not of the form $\cos\Omega$.

\subsection{Harmonic representation of the variations}
\label{sec-parallax-ba-attitude-harmonic}

From Eq.~(\ref{eq:delta_par}) it is seen that a global parallax shift
corresponds to variations of $\delta\Gamma$ and $\delta_z$
proportional to $\cos\Omega$ and $\sin\Omega$, respectively,
while $\delta_y$ and $\delta_x$ are constant. These quantities
correspond to terms of order $k=0$ and 1 in the more general
harmonic series
\begin{equation}\label{eq:har_gamma}
\left.
\begin{aligned}
\delta_x &=
\sum_{k\ge0}a_k^{(x)}\cos k\Omega+b_k^{(x)}\sin k\Omega\\
\delta_y &=
\sum_{k\ge0}a_k^{(y)}\cos k\Omega+b_k^{(y)}\sin k\Omega\\
\delta_z &=
\sum_{k\ge0}a_k^{(z)}\cos k\Omega+b_k^{(z)}\sin k\Omega\\
\delta\Gamma &=
\sum_{k\ge0}a_k^{(\Gamma)}\cos k\Omega+b_k^{(\Gamma)}\sin k\Omega\\
\end{aligned}
\quad\right\}\,.
\end{equation}
Specifically, if $a_k^{(\Gamma)}=b_k^{(\Gamma)}=0$ except for $a_1^{(\Gamma)}\ne 0$,
we find the following relations between the amplitude of the basic angle variation, the
global shift of the parallaxes, and the non-zero harmonics of the attitude errors:
\begin{equation}\label{eq:har_varpi}
\left.
\begin{aligned}
\delta\varpi &=\frac{1}{2R\sin\xi\sin(\Gamma_\mathrm{c}/2)}\,a_1^{(\Gamma)}
=0.8738\,a_1^{(\Gamma)}\\
a_0^{(y)} &=\frac{1}{\tan\xi\sin\Gamma_\mathrm{c}}\,a_1^{(\Gamma)}
=1.0429\,a_1^{(\Gamma)}\\
b_1^{(z)} &=\frac{1}{2\tan(\Gamma_\mathrm{c}/2)}\,a_1^{(\Gamma)}
=0.3734\,a_1^{(\Gamma)}\\
\end{aligned}
\quad\right\}\,.
\end{equation}
The numerical values correspond to the mean parameters relevant for \emph{Gaia},
that is $\Gamma_\mathrm{c}=106\fdg5$, $\xi=45\degr$, and $R=1.01$~au.

It is not the purpose of this paper to investigate the possible effects of
other harmonics of the basic-angle variation. However, it can be
mentioned that $a_1^{(\Gamma)}$ is the only harmonic parameter that is
degenerate with the attitude and stellar parameters. This means that
$a_0^{(\Gamma)}$, $b_1^{(\Gamma)}$, and $a_k^{(\Gamma)}$,
$b_k^{(\Gamma)}$ for $k>1$ can be determined as parameters in the
calibration model.
\section{Results of numerical simulations}
\label{s:simulations}

In this section we present the results of numerical tests performed
to check the above conclusions. To study different aspects of
the problem, we make use of two different, though complimentary,
solutions: a direct solution, where inversion of the normal matrix provides full
covariance information, and an iterative solution, with separate updates for
different groups of unknowns, similar to the method used in the actual
processing of \emph{Gaia} data \citep{2012A&A...538A..78L}.
While the direct solution can only handle a relatively small number of stars,
and therefore is of limited practical use, it enables us to investigate important
mathematical properties of the problem and to examine the correlations
between all the parameters. By contrast, the iterative solution cannot provide
this kind of information, but is more realistic in terms of the number of stars
and has been successfully employed in the processing of real \emph{Gaia}
data.

\subsection{Small-scale direct solutions}
\label{sec:small}

For the direct solutions a special simulation software was developed.
It simulates the observations of a small number of stars and the
reconstruction of their astrometric parameters based on conventional
least-squares fitting. The normal equations for the unknown parameters
are accumulated and the normal matrix is inverted using singular value
decomposition \citep{golub}. This decomposition allows us to study
mathematical properties of the problem, especially details of its degeneracy.
The simulations included $10^4$ stars uniformly distributed over the
celestial sphere. Observations were generated with a harmonic perturbation
of the basic angle as in Eq.~(\ref{eq:har_gamma}d) with
$a_1^{(\Gamma)}=1$~mas.
No noise was added to the observations in order to study the
effects of the basic angle variations in pure form. The solutions always
include five astrometric parameters per star: two components of the
position, the parallax, and two components of the proper motion.
Additional parameters representing the variations in attitude and basic
angle were introduced as required by various types of solutions.

The first test is to check the theoretical predictions in Eq.~(\ref{eq:har_varpi}).
To this end, a solution was made including only the star (S) and attitude (A)
parameters, where the latter were taken to be the harmonic amplitudes of
$\delta_x$, $\delta_y$, and $\delta_z$ as given by the first three equations
of (\ref{eq:har_gamma}) for $k\le 1$, i.e.\ with a total of nine attitude parameters.
The results of this solution, summarised in
Table~\ref{t_amplitudes}, are in very good agreement with the theoretical
expectations. Small deviations from the predicted values may be caused e.g.
by the limited number of stars, the finite size of the field of view, and
numerical rounding errors.

Another test is the singular value analysis of the normal matrix.
In addition to the solution described above (of type SA), we computed
three other solutions with different sets of unknowns but always using
the same observations. Solution S included only the star parameters as
unknowns, solution SB included also the harmonic coefficients of
$\delta\Gamma$ in the last equation of (\ref{eq:har_gamma}) for $k\le 1$,
and solution SBA included all three sets of unknowns.

The results, summarised in Table~\ref{t_sin_val}, again confirm
the theoretical predictions. The problem is close to being
degenerate only in case SBA where all three kinds of parameters
(star, basic angle, attitude) are fitted simultaneously. In this case
there is one singular value (${\sim}10^{-5}$) much smaller than
the second smallest value (${\sim}10^{-2}$). This indicates that the
problem has a rank deficiency of one, which is clearly caused by the
(near-)degeneracy between the global parallax shift and the specific
basic-angle and attitude variations described by Eq.~(\ref{eq:har_varpi}).
In all other cases (SB, SA, S) no isolated small singular values appear:
the problem is then formally well-conditioned, although (as we have seen
in case SA) the solutions may be biased by the basic-angle variations.

\begin{table}
\caption{The parallax shift and the attitude harmonics obtained in
the small-scale direct solution (of type SA) with $a_1^{(\Gamma)}=1$~mas
basic-angle variation.}
\label{t_amplitudes} 
\centering 
\begin{tabular}{ccc}
\hline\hline 
\noalign{\smallskip}
Quantity                     & Predicted      & Computed \\
                                  &  [mas]           &  [mas] \\
\noalign{\smallskip}
\hline
\noalign{\smallskip}
$\delta\varpi$              & 0.8738 & 0.8735  \\[5pt]
$a_0^{(x)}$ & 0 & $ 4\times 10^{-6}$ \\
$a_1^{(x)}$ & 0 & $ 9\times 10^{-6}$ \\
$b_1^{(x)}$ & 0 & $ 2\times 10^{-6}$ \\[7pt]
$a_0^{(y)}$ & 1.0429 & 1.0423 \\
$a_1^{(y)}$ & 0 & $ 2\times 10^{-5}$ \\
$b_1^{(y)}$ & 0 & $-2\times 10^{-6}$ \\[7pt]
$a_0^{(z)}$ & 0 & $-8\times 10^{-6}$ \\
$a_1^{(z)}$ & 0 & $-2\times 10^{-5}$ \\
$b_1^{(z)}$   & 0.3734 & 0.3733  \\

\noalign{\smallskip}
\hline 
\end{tabular}
\end{table}

\begin{table}
\caption{The singular values $\sigma_i$ of the normal matrix for the different
types of direct solutions.}
\label{t_sin_val}
\centering
\begin{tabular}{ccccc}
\hline\hline
\noalign{\smallskip}
 & SBA & SB & SA & S \\
\noalign{\smallskip}
\hline
\noalign{\smallskip}
$\sigma_1$            & $5.6\times 10^{-6}$ & 0.015  & 0.017  & 0.017 \\
$\sigma_2$            & 0.017               & 0.017  & 0.017  & 0.017 \\
\vdots                & \vdots              & \vdots & \vdots & \vdots \\
$\sigma_\mathrm{max}$ & 1549                & 387    & 1549   & 0.886 \\
\noalign{\smallskip}
\hline
\end{tabular}
\tablefoot{
The singular values are sorted lowest to highest. The different
solutions are denoted by the parameters included in the fit: S, B
and A stand respectively for the astrometric (star) parameters,
basic angle, and attitude angles.
}
\end{table}

The circumstance that the smallest singular value in case SBA is
not zero is partly attributable to rounding errors in a solution involving
$\ge 50\,000$ unknowns. However, even in exact arithmetics the
small-scale direct solution does not represent a fully degenerate
problem because
(i) the strict degeneracy only occurs if one neglects the finite size of the
fields of view, and
(ii) some parameters of the mission are slightly time-dependent, but
were assumed to be constant by the harmonic representations in
Eq.~(\ref{eq:har_gamma}). The question of the time dependence is
further addressed in Sect.~\ref{s:time}.

The attitude parameters used in the small-scale solutions are not representative
of any practically useful attitude model, but were chosen solely to verify
the expected degeneracy with the basic angle variation and parallax zero point.
In particular, the harmonic model of $\delta_x$, $\delta_y$, $\delta_z$
in Eq.~(\ref{eq:har_gamma}) cannot describe a solid rotation of the reference
frame, which explains why Table~\ref{t_sin_val} does not show the six-fold
degeneracy between the attitude and stellar parameters normally expected
from the unconstrained reference frame. This simplification is removed in the
large-scale simulations described below, which use a fully realistic attitude
model.

\subsection{Large-scale iterative solutions}

To test the effect of the basic angle variation in an iterative
solution, we make use of the \emph{Gaia} AGISLab simulation software
\citep[Appendix B]{2012A&A...543A..15H}. This tool allows us to simulate
independent astrometric solutions in a reasonable time, based on the
same principles as the astrometric global iterative solution
\citep[AGIS;][]{2012A&A...538A..78L} used for \emph{Gaia} but employing a
smaller number of primary stars and several other time-saving
simplifications.

To investigate the parallax zero point we have done a set of tests for different values of the basic angle in the range from $30\degr$ to $150\degr$, including the {\sc Hipparcos} and \emph{Gaia} values of $58\degr$ and $106\fdg5$, respectively. The nominal \emph{Gaia} scanning law and realistic geometry of the {\it Gaia}\/ fields of view are assumed in these simulations which include one million bright ($G=13$) stars uniformly distributed on the sky and observed during 5~years with no dead time. The nominal along-scan observation noise of 95~$\mu$as per CCD is assumed, based on the estimated centroiding performance of \emph{Gaia} for stars of $G=13$~mag. According to \citet{2012Ap&SS.341...31D} this corresponds to an expected end of mission precision of around 10~$\mu$as for the parallaxes. Additionally, basic-angle variations with an amplitude $a_1^{\left(\Gamma\right)} = 1$~mas are included. The unknowns consist of the five astrometric parameters per star and the attitude parameters based on a B-spline representation of the quaternion components \citep{2012A&A...538A..78L} using a knot interval of 30~s. The AGISLab simulations therefore incorporate many detailed features from the observations and data reductions of an actual scanning astrometric mission, including a number of higher-order effects neglected in our analytical approach.

\begin{figure}
 \resizebox{\hsize}{!}{\includegraphics{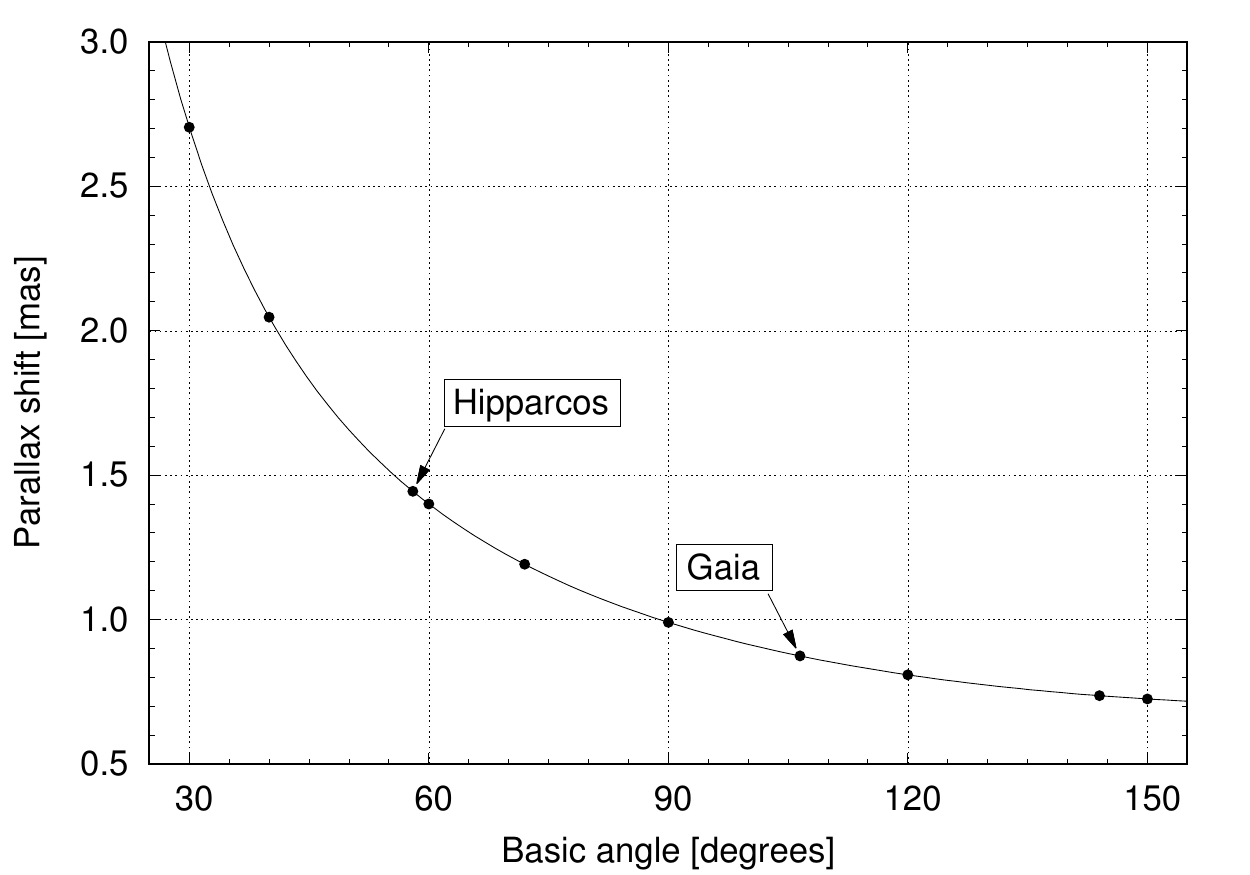}}
 \caption{Parallax zero point shift for a basic angle variation of the form $\cos\Omega$ with amplitude 1~mas. The dots show results of the large-scale iterative solution for several different basic angles, including the {\sc Hipparcos} and \emph{Gaia} values; the solid curve shows the theoretical relation from Eq.~(\ref{eq:har_varpi}).}
 \label{fig:parBA}
\end{figure}

The parallaxes obtained in the iterative solutions are offset from their "true" values (assumed in the simulation) by random and systematic errors. Results for the mean offset $\delta\varpi$ are shown by the points in Fig.~\ref{fig:parBA}. The curve is the theoretical relation from the first equation in (\ref{eq:har_varpi}). The points deviate from the curve by at most 1~$\mu$as, or $\lesssim0.1$\%. The random errors, measured by the sample standard deviation of the offsets of the individual parallaxes, were 7.3~$\mu$as in all the experiments, practically independent of the basic angle and in rough agreement with the expected end of mission precision.

\section{Discussion}
\label{sec:disc}

In the preceding sections it was shown that a global shift of the parallaxes is
observationally indistinguishable from a certain time-variation of the basic angle.
The relevant relation, strictly valid at the centre of the field of view, is given by
the last identity in Eq.~(\ref{eq:delta_par}). Here we proceed to discuss some practical
implications of this result.

\subsection{Physical relevance of the cos~$\Omega$ dependence}
\label{sec:relevance}

Elementary design principles have led to the choice of a nearly constant solar aspect angle $\xi$ (Fig.~\ref{fig:system}) for both {\sc Hipparcos} (43\degr) and \emph{Gaia} (45\degr). Moreover, for a satellite in orbit around the Earth or close to the second Lagrange (L$_2$) point of the Sun-Earth-Moon system, the barycentric distance $R$ is always close to 1~au. The form of basic-angle variation that is degenerate with parallax is then essentially proportional to $\cos\Omega$. This result is highly significant in relation to expected thermal variations of the instrument. The oblique solar illumination of the rotating satellite may produce basic-angle variations that are periodic with the spin period relative to the Sun, i.e.\ of the general harmonic form of the last line in Eq.~(\ref{eq:har_gamma}). A nearly constant, non-zero coefficient $a_1^{(\Gamma)}$ could therefore be a very realistic physical consequence of the way the satellite is operated.

Knowledge of the close coupling between a possible thermal impact on the instrument
and the parallax zero point has resulted in very strict engineering specifications for the
acceptable amplitude of short-term variations of the basic angle in both {\sc Hipparcos}
and \emph{Gaia}. In the case of \emph{Gaia}
it was known already at an early design phase that the basic angle variations cannot be fully avoided passively
and need to be measured. Therefore Gaia includes a dedicated laser-interferometric metrology
system, the basic angle monitor \citep[BAM;][]{2014SPIE.9143E..0XM}, to measure the short-term variations.
According to BAM measurements during the first year of the nominal operations of
\emph{Gaia}, the amplitude of the $\cos\Omega$ term, referred to 1.01~au and epoch
2015.0, was about 0.848~mas \citep{2016A&A...595A...4L}. Uncorrected, such a large
variation would lead to a parallax bias of 0.741~mas according to Eq.~(\ref{eq:har_varpi}).
For \emph{Gaia} Data Release~1 \citep{2016A&A...595A...2G} the observations were
corrected for the basic angle variations based on a harmonic model fitted to the
BAM measurements.

\subsection{Dependence on $\Gamma_\mathrm{c}$}
\label{s:smallGamma}

From Eq.~(\ref{eq:har_varpi}) it is seen that the parallax shift is inversely
proportional to $\sin(\Gamma_\mathrm{c}/2)$. For a constant
amplitude of the $\cos\Omega$ term in the basic angle variation, the parallax
shift therefore decreases with increasing basic angle (cf.\ Fig.~\ref{fig:parBA}).
From this point of view the optimum basic angle is therefore 180\degr. This
value would however be very bad for the overall conditioning and precision of
the astrometric solution
\citep{1997ESASP.402..823M,1998A&A...340..309M,2011EAS....45..109L},
which instead favours a value around 90\degr. Moreover, the sensitivity to the
$\cos\Omega$ variation is only a factor 1.4 smaller at 180{\degr} than at 90{\degr}.
The \emph{Gaia} value $\Gamma_\mathrm{c}=106\fdg5$ is therefore a reasonable
choice.

That the sensitivity to the basic angle variation increases with decreasing
$\Gamma_\mathrm{c}$ can be understood from simple arguments.
Consider the along-scan effects of the parallax shift $\delta\varpi$.
As long as the nominal basic angle $\Gamma_\mathrm{c}$ is large, the effects
in the two fields of view are significantly different. However, the smaller the basic
angle is, the more similar are the effects in the two fields of view. This can be
seen from Eq.~(\ref{eq:gh_par}) but is also obvious without any formula. We emphasise
that it is the basic angle variations that induce field angle perturbations and that the
solution tries to find parallaxes and attitude parameters that fit the perturbed field
angles. It is then obvious that a smaller basic angle will require a larger parallax shift
to absorb a basic-angle variation of given amplitude.

\subsection{Time dependence of $R$ and $\xi$}
\label{s:time}

As noted above, the barycentric distance $R$ and the angle $\xi$ are
not strictly constant but functions of time. In this case Eq.~(\ref{eq:delta_par})
gives the particular time dependences of $\delta\Gamma$, $\delta_x$,
$\delta_y$, and $\delta_z$ that are degenerate with a global parallax shift.
In particular,
\begin{equation}\label{eq:gamma_var}
\delta\Gamma(t) = C \, R(t)\sin\xi(t)\cos\Omega(t) \, ,
\end{equation}
where $C$ is constant, is indistinguishable from a parallax shift of
$\delta\varpi=\frac{1}{2}C/\sin(\Gamma_\mathrm{c}/2)$.

Any other form of the variation $\delta\Gamma(t)$ is not completely
degenerate with $\delta\varpi$ and may therefore contain components
that can be detected by analysis of the residuals and subsequently eliminated
by means of additional calibration terms. However, an arbitrary
variation $\delta\Gamma(t)$ in general also contains a component of the
form (\ref{eq:gamma_var}), which will result in some parallax shift.
This shift can be estimated by projecting the variation onto the function on
the right-hand side of Eq.~(\ref{eq:gamma_var}) in the least-squares sense:
\begin{equation}\label{eq:mean}
\delta\varpi={1\over 2\sin(\Gamma_\mathrm{c}/2)}\,
{
\bigl\langle\,\delta\Gamma(t)\,R(t)\,\sin\xi(t)\,\cos\Omega(t)\,\bigr\rangle
\over
\bigl\langle\,R(t)^2\,\sin^2\xi(t)\,\cos^2\Omega(t)\,\bigr\rangle
}\,,
\end{equation}
where the angular brackets denote averaging over time. If $R$ and $\xi$ are constant,
the factor $(R\sin\xi)^{-1}$ can be taken out of the averages. If, in addition, $\delta\Gamma(t)$
is strictly periodic in $\Omega$, we recover the first equality in Eq.~(\ref{eq:har_varpi}).

For Gaia, which operates in the vicinity of L$_2$, its
  barycentric distance $R$ varies between approximately 0.99 and 1.03~au
  as a combination of the eccentric heliocentric orbit of the Earth,
  the Lissajour orbit around L$_2$ and the time-dependent offset
  between the Sun and the Solar system's barycentre. As discussed in
  Sect.~\ref{sec:helio} below $\xi$ varies by about 1\% from its nominal
  value 45\degr.

\subsection{Effect of the finite size of the fields of view}
\label{sec:fov}

In order to derive Eqs.~(\ref{eq:gh_par}) and (\ref{eq:delta_par}) we neglected the finite
size of the field of view by considering only observations at the field centre ($g=h=0$).
This was necessary in order to obtain an exact relation between the parallax shift and
the basic-angle and attitude variations. In a finite field of view, additional terms appear
due to the variation of the parallax factor across the field, which cannot be represented
by a unique set of basic-angle variations. These terms are of the order of $\Phi\simeq 10^{-2}$
times smaller than the basic-angle variations, where $\Phi$ is half the size of the field of view.
As a consequence, a basic-angle variation of the form (\ref{eq:gamma_var}) is not strictly
degenerate with $\delta\varpi$ and the attitude angles when a finite field of view is considered.

However, if the instrument has also periodic optical distortions separately in each field of view
that need to be calibrated, the corresponding, more complex calibration model may contribute to the
degeneracy and, in worst case, restore complete degeneracy.

\subsection{Dependence on heliotropic coordinates}
\label{sec:helio}

The spherical coordinates $R$, $\xi$, $\Omega$ introduced in Sect.~\ref{sec-parallax} define
the position of the solar system barycentre in the scanning reference system (SRS). This is
what matters for calculating the parallax effect, which depends on the observer's displacement
from the barycentre. On the other hand, the physical relevance of the $\cos\Omega$ modulation
is connected with the changing illumination of the satellite by the Sun, which depends on the
heliocentric distance $R_\mathrm{h}$ and the (heliotropic) angles $\xi_\mathrm{h}$,
$\Omega_\mathrm{h}$ representing the proper direction towards the centre of the Sun at the
time of observation. The difference between the heliotropic and barytropic coordinates is at
most about 0.01~au and 0.01~rad, respectively. This is small but should be taken into account
for an accurate modelling of the basic-angle variations. In this context it can be noted
that the expected thermal impact on the satellite scales as $R_\mathrm{h}^{-2}$, while the
parallax factor scales as $R$.
On the other hand, the scanning law is chosen to keep $\xi_\mathrm{h}$ as constant as possible, while
$\xi$ can vary at the level of 1\%.
If the basic angle varies as
$R_\mathrm{h}^{-2}\sin\xi_\mathrm{h}\cos\Omega_\mathrm{h}$ it is no longer strictly
of the form (\ref{eq:gamma_var}). The resulting parallax bias can be estimated by means
of Eq.~(\ref{eq:mean}).

\subsection{Breaking the degeneracy?}
\label{sec:breaking}

If the basic angle varies as a consequence of the changing solar illumination of the rotating
satellite we expect to see a parallax bias according to Eq.~(\ref{eq:mean}). However, as discussed
above, the degeneracy with the parallax zero point is not perfect, and in principle this opens a
possibility to calibrate the basic angle variations from the observations. At least three different
effects that contribute to breaking the degeneracy could be used: the finite size of the field of
view (Sect.~\ref{sec:fov}), the time dependence of $R$ due to the eccentricity of the Earth's orbit
(Sect.~\ref{s:time}), and the difference between the barytropic and heliotropic angles
(Sect.~\ref{sec:helio}). Unfortunately all three effects only come in at a level of a few
per cent of the variation, or less, which makes the result very sensitive to small errors
in the calibration model. Moreover, the finite field of view is of little use if we have to calibrate
complex periodic variations of optical distortions independently in each field of view.
The best chance may be offered by the time
variation of $R$, where the ratio of the parallax effect to the illumination strength goes as
$R^3$ and consequently varies by $\pm 5$\% over the year. Thus the hope to break the
degeneracy purely from the observations themselves, i.e.\ based on the self-calibration principle,
is rather limited.

\section{Conclusions}
\label{s:conclusion}

We have presented an analysis of the effect of basic angle variations on
the global shift of parallaxes derived from observations by a scanning
astrometric satellite with two fields of view.

The method of small perturbations was used to derive the changes in the
four observables (the across- and along-field angles in both fields of
view) resulting from perturbations of four instrument parameters
(the basic angle and three components of the attitude). Conversely,
any given perturbation of the four observables can equally be represented
by a specific combination of the instrument parameters. Applying this
technique to the perturbations induced by a change in the parallax, we
derived the time-dependent variations of the instrument parameters that
exactly mimic a global shift of the parallaxes.

These relations confirm previous findings 
that an uncorrected variation of the basic angle of the form $a_1^{(\Gamma)}\cos\Omega$,
with $\Omega$ being the barycentric spin phase, leads to a global shift of the
parallax zero point of ${\simeq\,}0.87a_1^{(\Gamma)}$ for the parameters of the
\emph{Gaia} design. Results of numerical simulations are in complete agreement
with the analytical formulae.

In general, periodic variations of the basic angle can be expected from the thermal
impact of solar radiation on the spinning satellite \citep{lindegren77,LL:GAIA-LL-057}.
Those periodic variations are typically related to the heliotropic spin phase
$\Omega_\mathrm{h}$, which is close to the barycentric spin phase $\Omega$.
If the thermally-induced variations contain a significant component that is
proportional to $\cos\Omega_\mathrm{h}$, their effect on the observations
is practically indistinguishable from a global shift of the parallaxes. Although
the degeneracy is not perfect, it is difficult to break except by using of other
kinds of data or external information. In the case of \emph{Gaia} this includes, in
particular, direct measurement of the basic angle variations by means of laser
metrology (BAM). The use of astrophysical information such as parallaxes of
pulsating stars \citep{2011A&A...530A..76W,2016arXiv160900728G} and quasars
is vital for verifying the successful determination of the parallax zero point.

\begin{acknowledgements}
The authors acknowledge useful discussions with many colleagues within the \emph{Gaia} community, of which we especially wish to mention Ulrich Bastian, Anthony Brown, Jos de Bruijne, Uwe Lammers, Fran\c{c}ois Mignard, and Timo Prusti. The authors warmly thank the anonymous referee for valuable comments and suggestions. The work at Technische Universit\"at Dresden was partially supported by the BMWi grants 50\,QG\,0601, 50\,QG\,0901 and 50\,QG\,1402 awarded by the Deutsche Zentrum f\"ur Luft- und Raumfahrt e.V. (DLR).
\end{acknowledgements}

\bibliographystyle{aa}
\bibliography{parallaxShiftBA}

\end{document}